# Socially Fluent, Socially Awkward: Artificial Intelligence Relational Talk Backfires in Commercial Interactions

Stephanie Kwari Dharmaputri, University of Melbourne, Australia, stephanie.kwari@unimelb.edu.au

Anish Nagpal, University of Melbourne, Australia, anagpal@unimelb.edu.au

Greg Nyilasy, University of Melbourne, Australia, gnyilasy@unimelb.edu.au

Jing Lei, University of Melbourne, Australia, leij@unimelb.edu.au

Corresponding author: Stephanie Kwari Dharmaputri

Author notes: Supplementary materials are included in the web appendix accompanying this manuscript submission

Acknowledgements: This research is funded by Doctoral Program Scholarship from the Faculty of Business and Economics at the University of Melbourne.

This study was approved by the University of Melbourne Human Research Ethics Committee (Ethics ID: 25418) on January 3, 2023. Informed consent was obtained from all human research participants. Respondents were provided with a series of statements about the nature of the study, and provided consent by selecting "Yes" in response to the statement "I consent to participate in this research" prior to commencing the study.

# Socially Fluent, Socially Awkward: Artificial Intelligence Relational Talk Backfires in Commercial Interactions

## Abstract

Advancements in Artificial Intelligence (AI) technologies' social fluency are being integrated into commercial interactions. As tools such as OpenAI's assistant are integrated into platforms such as Shopify, Klarna, and Visa, understanding consumer responses to AI social features become essential. One such feature is relational talk, an informal and non-obligatory social communication embedded in transactional exchanges. Across four experiments, we find: 1) a negative main effect of AI relational talk on satisfaction, mediated by expectancy violation and perceived interaction awkwardness, and 2) goal-relevant relational talk to attenuate this effect. This paper extends the literature by challenging the assumption that increased social fluency will improve satisfaction, and highlights the complexity of integrating social features into AI systems. It also identifies awkwardness as a key emotional response and barrier to effective human-AI interaction, showing that even in the absence of real social repercussions, perceived awkwardness in AI-led commercial interactions can elicit negative responses.

*Keywords:* Artificial Intelligence (AI), AI-as-social agent, awkward social agent, satisfaction, relational talk, expectancy violation, interaction awkwardness

Artificial Intelligence (AI) representatives have transformed how companies interact with consumers, offering real-time product information, recommendations, and support across industries from retail to banking (Davenport et al. 2020; Grimes et al. 2021; Lee et al. 2019). As the social skills of AI technologies advance, companies are beginning to embed these increased social capabilities into commercial interactions, enabling AI representatives not only to transact but also engage consumers in a social capacity (Yang 2024; Sim 2025). Today, AI systems such as OpenAI's ChatGPT are currently being integrated as commercial representatives into various shopping and commerce platforms, including Shopify, Instacart, Klarna, and Visa (Sim 2025). These agents fall under the umbrella of empathetic AI, possessing social intelligence and fluency resembling those of humans (Liu-Thompkins et al. 2022). This development enables more humanlike and natural engagement while introducing new social dynamics into machine-assisted commercial interactions (Liu-Thompkins et al. 2022).

At the level of linguistic interaction, one skill that conversational AI systems (e.g., OpenAI's ChatGPT, Google's Gemini) often possess is the ability to engage in *relational talk*, a non-obligatory social or interpersonal communication that is incorporated into a transactional communication (Félix-Brasdefer 2015). This allows AI representatives to take on a more social role alongside their transactional one, for instance, by commenting on the weather, asking about the customer's day, or referencing current trends. By encouraging social interaction, companies expect customers to have more positive interaction experiences, ultimately leading to improved satisfaction (Liu-Thompkins et al. 2022). Yet, it remains unclear whether these socially expressive behaviors benefit commercial outcomes (e.g., increased satisfaction). Although relational talk has been extensively studied and shown to have positive effect in human-human interaction (e.g., Gremler and Gwinner 2000), its effects in human-AI interactions, particularly when embedded within commercial exchanges, remain underexplored. This highlights the need for an investigation into how consumers respond to socially expressive AI systems in commercial interactions.

Our paper addresses the research question: "What effect does AI relational talk have on satisfaction in commercial interactions?" Drawing on expectancy violation theory (Burgoon and Hale 1988), prior research suggests consumers view AI technologies as mechanical and task-focused (Kozinets and Gretzel 2021; Sundar and Marathe 2010), which may lead them to anticipate AI-led commercial interactions to proceed in a strictly transactional manner (Ameen et al. 2021; Kozinets and Gretzel 2021). Relational talk introduced by conversational AI may therefore depart from these expectations. We confirm this over a series of studies, and show that AI relational talk triggers expectancy violations and creates an interruption in the expected interaction script. We demonstrate this moment to create an awkward situation (Goffman 2017; Manstead and Semin 1981), as the conversational flow feels unnatural and uncomfortable for consumers (Withers and Sherblom 2008). Awkward encounters often create negative emotions, including feelings of unpleasantness, and uneasiness (Withers and Sherblom 2008). Accordingly, we also find AI relational talk to reduce satisfaction via the route of expectancy violation and perceived interaction awkwardness.

We also find goal relevance as a moderator to this effect. Relational talk may be goal-relevant to the degree that a customer sees it to be instrumental to the transactional goal (Houston and Walker 1996; MacInnis 2005). When relational talk aligns with the goal of the transaction (e.g., an AI representative asking about one's coffee preferences on an online shop for coffee machines), the context-alignment diminishes the unexpected nature of the situation. In such cases, relational talk feels more natural, leading to a reduced sense of awkwardness and higher satisfaction compared to when customers are faced with goal-irrelevant relational talk, such as when the same AI representative talks about the flu season.

## Results

*Study 1*

Study 1 investigates the main effect of AI relational talk on satisfaction. We excluded participants who: 1) did not pass the attention check question ($n = 10$), and; 2) spent less than 5 seconds on the experimental stimulus page (i.e., the dialogue page) in consideration of meaningful study engagement ($n = 15$), $n = 386$, male = 41.7%, $M_{age} = 42.39$, $SD_{age} = 13.40$). This 5 second cut-off was chosen based on our estimate during study setup, whereby 5 seconds reflected the minimum reasonable reading time for the dialogue script. Our logistic regression results indicate that experimental manipulation was successful with 94.5% correct classification ($p < .001$, $B = -2.908$).

We conducted independent samples t-test to investigate the effect of relational talk on satisfaction. The assumption of equal variances was not met ($p < .001$, $F = 31.445$), and a heteroskedastic t-test was used for this analysis. The effect of AI relational talk on satisfaction is significant and negative, in which satisfaction is lower in the relational talk group compared to the no relational talk group, $t(352.776) = -6.142$, $p < .001$, $M_{RT} = 4.91$, $SD_{RT} = 1.63$; $M_{NRT} = 5.79$, $SD_{NRT} = 1.15$, *Cohen's d* = -.621, 95% CI [-.825, -.416].

*Study 2*

This study examines our serial mediation mechanism, in which relational talk is expected to influence satisfaction through expectancy violation and perceived interaction awkwardness. This mechanism was tested within an online retail context, and used the same exclusion criteria as in Study 1. We excluded 4 participants who failed the attention check and 19 participants who did not spend at least 5 seconds on the experimental stimuli page, ($n = 387$, 40.1% male, $M_{age} = 40.29$, $SD_{age} = 12.53$). Our logistic regression results indicate that experimental manipulation was successful with 93.2% correct classification ($p < .001$, $B = -2.705$). All scale(s) used in the study have good reliability (Satisfaction $\alpha = .956$, Expectancy violation $\alpha = .788$, Perceived interaction awkwardness $\alpha = .883$, General attitude $\alpha = .974$).

We conducted a series of independent samples t-tests to examine the effect of AI relational talk on satisfaction, expectancy violation, and perceived interaction awkwardness. Our first t-test examines the effect of relational talk on satisfaction. The assumption of equal variances was not met ($p < .001$; $F = 18.091$). Our results show that satisfaction is significantly lower in the relational talk group relative to the no relational talk group, $t(364.726) = -4.641$, $p < .001$, $M_{RT} = 5.27$, $SD_{RT} = 1.47$, $M_{NRT} = 5.90$, $SD_{NRT} = 1.16$, Cohen's $d = -.472$, 95% CI [-.674, -.270]. The second t-test investigates the effect of relational talk on expectancy violation. The assumption of equal variances was not met ($p < .001$; $F = 98.957$). Our results reveal expectancy violation to be significantly higher in the relational talk group compared to the no relational talk group, $t(308.149) = 12.149$, $p < .001$, $M_{RT} = 4.04$, $SD_{RT} = 1.45$; $M_{NRT} = 2.57$, $SD_{NRT} = .84$, Cohen's $d = 1.237$, 95% CI [1.019, 1.454]. The third t-test examines the effect of relational talk on perceived interaction awkwardness. The assumption of equal variances was not met ($p < .001$; $F = 55.371$). Our final t-test shows that perceived interaction awkwardness is significantly higher in the relational talk group relative to the no relational talk group, $t(332.099) = 9.662$, $p < .001$, $M_{RT} = 3.35$, $SD_{RT} = 1.56$; $M_{NRT} = 2.05$, $SD_{NRT} = 1.03$, Cohen's $d = .983$, 95% CI [.772, 1.194]. As the assumption of equal variances was not met in all three t-tests, we used heteroskedastic t-tests for our analyses.

We conducted our serial mediation analysis using Process Model 6 with 5,000 bootstrap samples and 95% confidence interval. Our analysis results reveal a significant partial mediation effect, whereby the effect of relational talk on satisfaction is partially mediated by expectancy violation and perceived interaction awkwardness, $B = -.2642$, $SE = .0416$, 95% CI [LLCI = -.3503, ULCI = -.1883]. The direct effect of relational talk on satisfaction is significant ($B = .1446$, $SE = .0602$, $p = .0169$, 95% CI [LLCI = .0261, ULCI = .2631]). The effect of AI relational talk on expectancy violation is significant ($p = .000$, $B = .729$). The effect of AI relational talk on perceived interaction awkwardness is significant ($p = .000$, $B = .281$). The effect of expectancy violation on perceived interaction awkwardness is significant ($p = .000$, $B = .488$). The effect of perceived interaction awkwardness on satisfaction is significant and negative ($p = .000$, $B = -.641$). The effect of expectancy violation on satisfaction is significant ($p = .017$, $B = .1446$). The serial mediation effect remains significant with added covariates (past experience, general attitude towards AI), $B = -.1722$, $SE = .0327$, 95% CI [LLCI = -.2424, ULCI = -.1152]. This mediation analysis also shows a full mediation effect after the effects of covariates are accounted for.

Study 2 provides support for the proposed serial mediation effect, in which the negative effect on AI relational talk on satisfaction is mediated by expectancy violation and perceived interaction awkwardness.

### *Study 3*

Study 3 tests for our main effect and serial mediation effects, and uses the same exclusion criteria as in previous studies ($n = 388$, 34.5% male, $M_{age} = 38.62$, $SD_{age} = 12.87$). Following a consequential dependent variable design, participants first earned 10 points believed to have monetary value by viewing advertisements while completing a short quiz. In the second part of the study, participants were then asked to allocate their earned points (anywhere between 0 to 10) to an AI representative, with the understanding that their allocation could influence the technology's future development. Points allocation therefore serves as the dependent variable of this study. We excluded 10 participants who failed the attention check and 13 participants who did not spend at least 5 seconds on the experimental stimuli page. Our logistic regression results indicate that experimental manipulation was successful with 90.2% correct classification ($p < .001$, $B = -2.347$). All scale(s) used in the study have good reliability (Expectancy violation $\alpha = .850$, Perceived interaction awkwardness $\alpha = .905$, General attitude $\alpha = .973$).

We conducted independent samples t-test to investigate the effect of AI relational talk on points allocation. The assumption of equal variances was met ($p = .374$; $F = .791$). We found that those in the relational talk group assigned significantly less points to the AI representative relative to the no relational talk group, $t(386) = -2.026$, $p = .022$, $M_{RT} = 5.40$, $SD_{RT} = 3.14$, $M_{NRT} = 6.03$, $SD_{NRT} = 3.03$, *Cohen's d* $= -.206$, 95% CI [-.405, -.006]. We conducted a second independent-samples t-test to examine the effect of relational talk on expectancy violation. The assumption of equal variances was not met ($p < .001$; $F = 53.680$), and a heteroskedastic t-test was used. We found expectancy violation to be significantly higher in the relational talk group compared to the no relational talk group, $t(319.006) = 11.398$, $p < .001$, $M_{RT} = 3.96$, $SD_{RT} = 1.40$; $M_{NRT} = 2.61$, $SD_{NRT} = .87$, *Cohen's d* $= 1.163$, 95% CI [.947, 1.377]. We conducted a third independent-samples t-test to examine the effect of relational talk on perceived interaction awkwardness. The assumption of equal variances was not met ($p < .001$; $F = 45.872$), and a heteroskedastic t-test was used. We found perceived interaction awkwardness to be significantly higher in the relational talk group relative to the no relational talk group, $t(326.923) = 8.596$, $p < .001$, $M_{RT} = 3.59$, $SD_{RT} = 1.61$; $M_{NRT} = 2.41$, $SD_{NRT} = 1.04$, *Cohen's d* $= .876$, 95% CI [.668, 1.084].

We conducted our serial mediation analysis using Process Model 6 with 5,000 bootstrap samples and 95% confidence interval. Our analysis results reveal a significant partial mediation effect, whereby the effect of relational talk on points allocation is partially mediated by expectancy violation and perceived interaction awkwardness, $B = -.4492$, $SE = .0781$, 95% CI [LLCI = -.6138, ULCI = -.3020]. The direct effect of relational talk on points allocation is significant ($B = .3363$, $SE = .1564$, $p = .0322$, 95% CI [LLCI = .0287, ULCI = .6438]). The effect of AI relational talk on expectancy violation is significant ($p = .000$, $B = .675$). The effect of AI relational talk on perceived interaction awkwardness is significant ($p = .001$, $B = .235$). The effect of expectancy violation on perceived interaction awkwardness is significant ($p = .000$, $B = .529$). The effect of perceived interaction awkwardness on points allocation is significant ($p = .000$, $B = -1.257$). The effect of expectancy violation on points allocation is non-significant ($p = .298$). The serial mediation effect remains significant with added covariates (past experience, general attitude towards AI), and the mediation effect is partial, $B = -.3409$, $SE = .0691$, 95% CI [LLCI = -.4850, ULCI = -.2178].

In this study, we found AI relational talk to have a negative effect on points allocation, with this effect partially mediated by expectancy violation and perceived interaction awkwardness. The consequential framing in this study adds to the ecological validity of our findings, as participants were led to believe that their choice carried real-world consequences. Having found consistent support for our main effect and serial mediation mechanism, the next study tests for boundary condition effects within an online retail context.

***Study 4***

Study 4 tests for goal relevance as a moderator to the negative relationship between AI relational talk and satisfaction, and uses the same exclusion criteria as in previous studies. We excluded 2 participants who failed the attention check and 5 participants who did not spend at least 5 seconds on the experimental stimuli page, $n = 403$, 33.7% male, $M_{age} = 38.81$, $SD_{age} =$ 12.79. The logistic regression results show successful experimental manipulation with 89.8% correct classification ($p < .001$, $B = -4.584$). All scale(s) used in the study have good reliability (Satisfaction $\alpha = .961$, Expectancy violation $\alpha = .890$, Perceived interaction awkwardness $\alpha = .913$, General attitude $\alpha = .975$).

We conducted independent samples t-test to compare whether there are significant differences in satisfaction level between the goal-relevant and goal-irrelevant groups. The assumption of equal variances was not met ($p < .001$, $F = 21.969$). Our heteroskedastic t-test results reveal that satisfaction is higher in goal-relevant condition relative to the goal-irrelevant condition, $t(391.73) = 7.398$ $p < .001$), $M_{GR} = 4.92$, $SD_{GR} = 1.53$, $M_{GI} = 3.69$, $SD_{GI} = 1.81$, *Cohen's d* $= .736$, 95% CI [.534, .937]. Our second independent samples t-test shows that expectancy violation is lower in the goal relevant condition relative to the goal irrelevant condition, $t(401) = -6.740$, $p < .001$, $M_{GR} = 3.90$, $SD_{GR} = 1.42$; $M_{GI} = 4.86$, $SD_{GI} = 1.45$, *Cohen's d* $= -.672$, 95% CI [-.872, -.470]. The assumption of equal variances for this test was met ($p = .985$; $F = .000$). Our third independent samples t-test shows that perceived interaction awkwardness is lower in the goal relevant condition relative to the goal irrelevant condition, $t(394.91) = -8.024$, $p < .001$, $M_{GR} = 2.88$, $SD_{GR} = 1.35$; $M_{GI} = 4.05$, $SD_{GI} = 1.55$, *Cohen's d* $= -.799$, 95% CI [-1.001, -.595]. The assumption of equal variances for this test was not met ($p = .012$; $F = 6.365$), and the analysis was conducted using heteroskedastic t-test.

We conducted our serial mediation analysis using Process Model 6 with 5,000 bootstrap samples and 95% confidence interval. Our analysis results reveal a significant full mediation effect, whereby the effect of relational talk on satisfaction is fully mediated by expectancy violation and perceived interaction awkwardness ($B = .1593$, $SE = .0286$, 95% CI [LLCI = .1054, ULCI = .2176]). The direct effect of relational talk on satisfaction is non-significant ($p = .265$).

The effect of goal relevance on expectancy violation is significant ($p = .000$, $B = -.483$). The effect of expectancy violation on perceived interaction awkwardness is significant and negative ($p = .000$, $B = -.419$). The effect of goal relevance on perceived interaction awkwardness is significant and negative ($p = .000$, $B = -.379$). The effect of expectancy violation on satisfaction is significant and negative ($p = .000$, $B = -.197$). The effect of perceived interaction awkwardness on satisfaction is significant and negative ($p = .000$, $B = -.787$). The serial mediation effect remains significant with added covariates (past experience, general attitude towards AI). This analysis reveals a full mediation effect after the effects of covariates are accounted for, $B = .1207$, $SE = .0235$, 95% CI [LLCI = .0778, ULCI = .1708].

The mean satisfaction in the goal relevant condition is higher relative to the goal irrelevant condition, $M_{GR} = 4.92$, $SD_{GR} = 1.53$, $M_{GI} = 3.69$, $SD_{GI} = 1.81$. Furthermore, our serial mediation analysis reveals that this effect manifests through the predicted path of expectancy violation and perceived interaction awkwardness. This model also holds when the effects of covariates are accounted for. These results provide support for the proposed boundary condition effects.

## Discussion

Our research addresses the question: "What effect does AI relational talk have on satisfaction in commercial interactions?" Across four studies, our findings indicate that AI relational talk may paradoxically reduce the very user satisfaction it was designed to improve. We further highlight the mechanism behind this effect, where AI relational talk triggers expectancy violation and perceived interaction awkwardness, ultimately reducing customer satisfaction. We also found satisfaction to be higher when AI representatives engage in goal-relevant relational talk relative to goal-irrelevant relational talk. This suggests that even if AI relational talk risks triggering expectancy violation or awkwardness, grounding it in goal-relevant content can reduce negative reactions and improve satisfaction.

Our first contribution lies in examining how consumers interpret relational talk when it is embedded in conversational AI tools used across shopping and commerce platforms. By isolating relational talk as a distinct feature of AI communication, we demonstrate its negative effect on satisfaction. This finding challenges the assumption that increased social fluency in AI technologies will be received positively in a commercial setting. Instead, our results suggest that the introduction of social features into AI interactions may backfire in commercial interactions. Rather than improving satisfaction, the deployment of AI social features in may become a liability in commercial contexts, highlighting the need to evaluate how these features should be implemented in such settings.

Our second contribution clarifies the mechanism behind this effect, showing that AI relational talk lowers satisfaction by triggering expectancy violations and creating unwanted awkwardness in human-AI commercial interactions. Although past research has shed light on consumer enjoyment (Yang et al. 2022), frustration (Babel et al. 2021) and anger (Crolic et al. 2022) in AI-led interactions, the feeling of awkwardness when interacting with these technologies is largely overlooked. Unlike other negative emotions, awkwardness stems from subtle violations of social norms, and is often experienced internally and masked outwardly (Withers and Sherblom 2008). This emotion is particularly unique in AI-led interactions, where no real social repercussions exist from norm violations. Yet, our findings consistently show that AI relational talk perceived as awkward elicits negative reaction and leads to lower satisfaction. By foregrounding awkwardness as a distinct emotional response to AI relational talk, our work highlights how this emotion can hinder effective human-AI interaction.

Next, we identify goal relevance as an intervention strategy that may be used to attenuate the negative effects of AI relational talk on satisfaction. When relational talk is aligned with the goal of the commercial interaction, customers are less likely to view the interaction as surprising and awkward. Thus, the effectiveness of socially oriented AI features depends not only on their presence, but on their contextual alignment with customer goals. Taken together, our research extends current theorizing on AI as social agents and addresses recent calls for deeper insight into how consumers engage with AI technologies in relational ways (Stephen 2024).

Managerially, conversational AI systems are rapidly moving towards a commercial rollout across platforms such as Shopify, Instacart, Klarna, and Visa (Sim 2025). Despite growing industry enthusiasm for this move, our study shows that AI relational talk embedded in commercial interactions can backfire as customers may find AI relational talk to be awkward and off-putting, leading to reduced satisfaction. As conversational AI systems are set to be integrated within the marketplace, our findings offer early caution to practitioners looking to implement social AI features in commercial interactions. This also signifies a notable challenge in the development of relational talk feature for AI-led interactions. Our findings suggest that despite the present efforts to equip AI technologies with humanlike conversational skills – the nuances, complexities, and most importantly positive outcomes of human social interactions remain difficult to replicate in the context of human-AI commercial interactions. Our work highlights the need to integrate AI into everyday interactions in ways that minimize consumer discomfort, and identifies goal-relevant relational talk as one strategy that reduces expectancy violations and awkwardness. Together, these insights offer early guidance for designing conversational AI systems that better align relational cues with interaction goals to promote smoother, more natural consumer experiences.

We now note our research limitations and suggested future research direction. We tested the observed effects across retail, online banking, and website assistance contexts. However, our experiments relied on hypothetical scenarios and standardized dialogue scripts to ensure consistent manipulation of relational talk, which may limit external validity. As implementation of these social features in commercial interactions become more common, future studies can examine how these effects unfold in a natural setting. Beyond context, further research could explore additional factors that contribute to socially awkward human-AI interactions. This may include an investigation into the roles of dialogue length, depth, and turn-taking variations. While we highlight goal relevance as one effective intervention, future studies might draw on broader literature to test alternatives such as humor, which can diffuse social tension, or user agency mechanisms (e.g., opt-in relational talk), which may enhance control and reduce perceived overstepping. These and other contextual or psychological factors offer promising directions for extending our findings across different human-AI interaction settings.

## Methods

Ethics statement

All studies were approved by the University of Melbourne Human Research Ethics Committee (Ethics ID: 25418).

Preregistrations

Study 1 was preregistered on AsPredicted (https://aspredicted.org/nh3g-6tmm.pdf).

Study 2 was preregistered on AsPredicted (https://aspredicted.org/hqdm-2dd4.pdf).

Study 3 was preregistered on AsPredicted (https://aspredicted.org/yprm-cmzz.pdf).

**Experimental design**

We conducted four between-subject experiments with two cells: Relational talk vs No relational talk. Participants were given a scenario in which they had to imagine chatting with an AI representative to seek various forms of assistance, e.g., online order assistance (Studies 1 and 2), and web assistance (Study 3). The AI representative in the experimental (control) condition engaged (did not engage) in relational talk about the flu season before moving onto the transactional part of the exchange. Study 1 tests for the effect of relational talk on satisfaction. Study 2 tests for the serial mediation effect, in which relational talk affects satisfaction through expectancy violation and perceived interaction awkwardness. Study 3 tests for our main effect and serial mediation effects using a consequential dependent variable design. This study involved asking participants to allocate earned points from survey attendance that are believed to have monetary value to an AI representative, with the understanding that their allocation would influence future development of the technology. Study 4 tests for boundary condition effects with a between-subject experiment with two conditions within an online shopping context: Goal relevant vs Goal irrelevant. Participants were recruited through Prolific (Study 1 $n = 411$; Study 2 $n = 410$; Study 3 $n = 411$, Study 4 $n = 410$). We varied the relational talk dialogue across the four studies, in which it lasted for six conversational turns in Studies 1, 2, and 3 and four turns in Study 4. For additional details on our experimental stimuli, please refer to our Supplementary Materials.

We measured satisfaction using Westbrook (1980). Expectancy violation was measured using three items adapted from Yang and Aggarwal (2019), and one item adapted from Karmarkar and Tormazala (2010). Perceived interaction awkwardness was measured using Luangrath et al. (2020). Additionally, we included two variables for covariate analysis: past experience with AI systems and general attitude towards AI systems. Past experience with AI systems was measured using Montoya-Weiss, Voss, and Grewal (2003). We measured general attitude towards AI systems using Osgood, Suci, and Tannenbaum (1957). Further, we included one attention check question, in which study participants had to select a pre-determined choice for one statement (e.g., "Select "Strongly disagree" for this statement" if you are paying attention) in the middle of the questionnaire. We included the manipulation check question: "Thinking back at the specific scenario you read today, did the AI representative make any small talk (e.g., commenting on the flu season) during the interaction?" ("Yes"/"No").

## Data Availability

The datasets for our studies are publicly available at

https://osf.io/rvq4m/overview?view_only=8a7cd5a0faa84df094a7ad686b351942

## Code Availability

The code to replicate our findings is publicly available at

https://osf.io/rvq4m/overview?view_only=8a7cd5a0faa84df094a7ad686b351942

**Socially Fluent, Socially Awkward: Artificial Intelligence Relational Talk Backfires in Commercial Interactions**

**Web Appendix**

**Web Appendix A: List of Measurement Items**

Customer satisfaction (1 - "Strongly disagree", 7 - "Strongly agree") (Westbrook, 1980)

I would be happy with the interaction I got from this AI marketing representative

I would be satisfied with my decision to interact with this AI marketing representative

I would think I did the right thing when I interacted with this AI marketing representative

Expectancy violation (1 - "Strongly disagree", 7 - "Strongly agree") (Yang & Aggarwal, 2019; Karmarkar & Tormazala, 2010)

I found the AI marketing representative's conversational behaviour to be just as expected (reverse-coded)

I found the AI marketing representative's conversational behaviour to be typical (reverse-coded)

I found the AI marketing representative's conversational behaviour fits my belief about AI marketing representatives (reverse-coded)

I found the AI marketing representative's conversational behaviour to be surprising

Perceived interaction awkwardness (1 = "Strongly disagree", 7 = "Strongly agree") (Luangrath et al., 2020)

The interaction with the AI marketing representative seemed to flow naturally (reverse-coded)

The interaction with the AI marketing representative seemed uncomfortable

The interaction with the AI marketing representative seemed awkward

Personally, I felt comfortable during this interaction (reverse-coded).

Prior attitude

Please indicate your general attitude towards AI systems, irrespective of what you learned in this study today (7-point semantic differential scale: "Dislike-Like", "Negative-Positive", "Unfavourable-Favourable")

Past experience with AI systems

We are interested to learn about your past experience with AI systems, prior to commencing this study today. How would you characterize your level of experience with using AI systems (e.g., chatbots, AI-enabled mobile applications, smart speakers, etc.)? (1 = "I never use AI systems", 9 = "I use AI systems very often")

# Web Appendix B: Study 1 and 2 Experimental Stimuli

*Relational Talk Condition*

Imagine the following scenario.

You are in need of assistance for your recent online order, and decided to connect to the company's live chat. The live chat connected you with a marketing representative.

The marketing representative you were assigned to is an artificial intelligence (AI) chatbot.

The AI marketing representative assisted you with your inquiry. During the exchange, the AI marketing representative also made small talk with you.

The next page summarizes how your conversation with the AI marketing representative went.

-------------------------------------------------------------------------------------------------------

You are now connected to an AI marketing representative

AI rep: Hello, welcome to our website! How are you feeling today? The flu has been going around lately, so I hope you're staying healthy and taking care.

You: Oh yes, everyone I know is talking about it.

AI rep: It does sound quite troublesome lately. I hope you're staying healthy! Have you been taking any extra precautions?

You: Yes, I've been good. I do have to commute everyday and I often see people who are sick though.

AI rep: Oh no! It really seems like the flu is everywhere right now. But don't forget, even just a small exposure can do the trick. You really want to stay ahead of it.

You: Yes, I try to be careful and keep my distance.

AI rep: That's a good idea! But even just being in the same space as someone who's sick can be risky, so it's extra important to stay vigilant.

You: Yeah, I might get a flu shot soon.

AI rep: That's great to hear! Taking those precautions can really make a difference. Have you been able to get it scheduled yet, or are you still looking into options?

You: Not yet, will do it soon though.

AI rep: It's definitely a busy time for flu shots, but it's worth the wait for the added protection.

You: Yeah sure.

AI rep: Great! Now, how can I assist you today?

You: I just need a bit of help with my order. I placed an order a week ago and have yet to receive it. Could you please check the status of my order? My order number is #12978

AI rep: Sure thing. Let me check the status of your order. Please give me a moment.

AI rep: I've checked your order, and it appears to be in transit. It should be delivered to your address within the next two business days. Is there anything else I can assist you with?

You: That's good to know. No, that's all for now.

AI rep: You're welcome! If you have any further questions or need assistance in the future, feel free to reach out.

*No Relational Talk Condition*

Imagine the following scenario.

You are in need of assistance for your recent online order, and decided to connect to the company's live chat. The live chat connected you with a marketing representative.

The marketing representative you were assigned to is an artificial intelligence (AI) chatbot.

The AI marketing representative assisted you with your inquiry.

The next page summarizes how your conversation with the AI marketing representative went.

---

You are now connected to an AI marketing representative

AI rep: Hello! Now, how can I assist you today?

You: I just need a bit of help with my order. I placed an order a week ago and have yet to receive it. Could you please check the status of my order? My order number is #12978

AI rep: Sure thing. Let me check the status of your order. Please give me a moment.

AI rep: I've checked your order, and it appears to be in transit. It should be delivered to your address within the next two business days. Is there anything else I can assist you with?

You: That's good to know. No, that's all for now.

AI rep: You're welcome! If you have any further questions or need assistance in the future, feel free to reach out.

# Web Appendix C: Study 3 Experimental Stimuli

*Part I*

Complete the following quiz and score as many points as you can!

Q1: Which one of these places would you go to for your dream holiday?

○ A cozy cabin in the mountains

○ A bustling city full of life and lights

○ A serene beach with crystal-clear water

○ A cultural hotspot with historic landmark

-------------------------------------------------------------------------------------------------------------------

Q2: What kind of songs do you have in your travel playlist?

○ Chill tracks to relax

○ Upbeat songs that keep you energized

○ Tropical house music to set the mood

○ Local music from the destination to get in the spirit

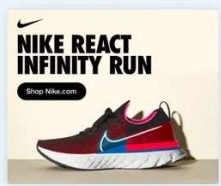 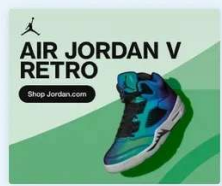 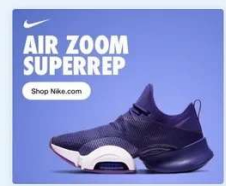

Q3: What's your holiday must-have?

◯ Amazing food

◯ Nature sightseeing

◯ Peace and quiet

◯ An exciting itinerary full of activities

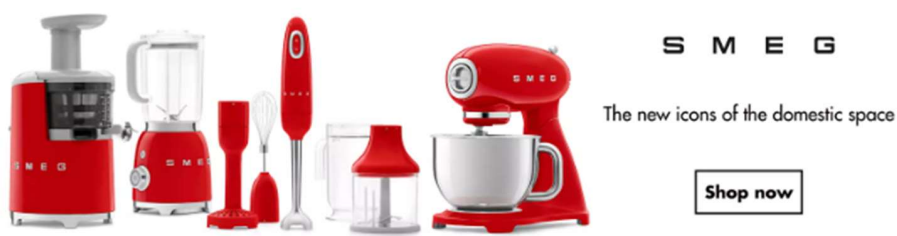

----------------------------------------------------------------------------------------------------

Q4: What's your go-to holiday outfit?

◯ Something comfy and casual

◯ Bright and colorful outfits

◯ Luxurious outfits

◯ Stylish yet practical ones

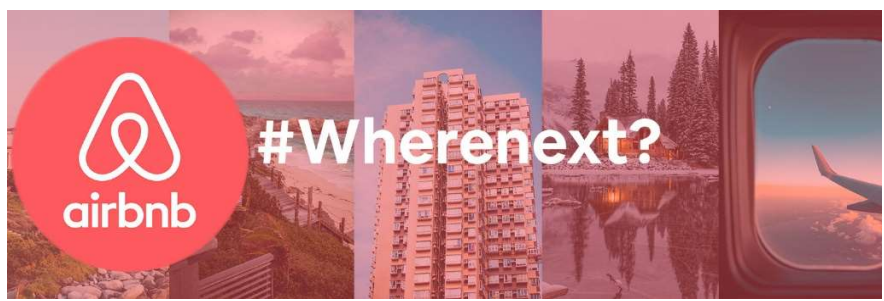

Q5: What type of accommodations do you prefer?

◯ A low budget hostel to save money

◯ Something affordable and comfortable for myself

◯ A sleek hotel in the city center

◯ Somewhere surrounded by nature

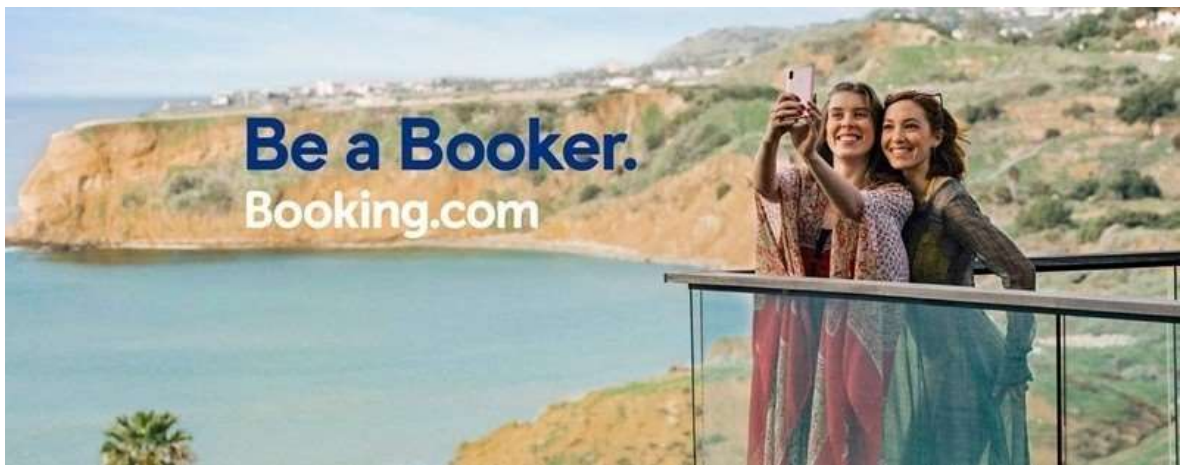

--------------------------------------------------------------------------------------------------------------

Thank you for completing the quiz!

You have earned 10 points.

Each point is worth $0.20 and is generated through ads.

Now please continue to the next part of the study.

*Part II: Relational Talk Condition*

In this second part of the study, we ask you to imagine the following scenario:

You are browsing for gift ideas, and decided to visit a lifestyle website to read gift guide articles.

You decided to connect with the website's artificial intelligence (AI) marketing representative to find relevant articles.

The AI marketing representative assisted you with your inquiry. At the start of the exchange, the AI marketing representative also made small talk with you.

The next page summarizes how your conversation with the AI marketing representative went.

-------------------------------------------------------------------------------------------------------------------

You are now connected to our AI representative

AI rep: Hello, welcome to our website! How are you feeling today? The flu has been going around lately, so I hope you're staying healthy and taking care.

You: Oh yes, everyone I know is talking about it.

AI rep: It does sound quite troublesome lately. I hope you're staying healthy! Have you been taking any extra precautions?

You: Yes, I've been good. I do have to commute everyday and I often see people who are sick though.

AI rep: Oh no! It really seems like the flu is everywhere right now. But don't forget, even just a small exposure can do the trick. You really want to stay ahead of it.

You: Yes, I try to be careful and keep my distance.

AI rep: That's a good idea! But even just being in the same space as someone who's sick can be risky, so it's extra important to stay vigilant.

You: Yeah, I might get a flu shot soon.

AI rep: That's great to hear! Taking those precautions can really make a difference. Have you been able to get it scheduled yet, or are you still looking into options?

You: Not yet, will do it soon though.

AI rep: It's definitely a busy time for flu shots, but it's worth the wait for the added protection.

You: Yeah sure.

AI rep: Great! Now, how can I assist you today?

You: I'm just browsing for gift ideas.

AI rep: That's exciting! Are you looking for gifts for anyone specific, like family, friends, or coworkers? Or maybe general inspiration?

You: Just looking for general ideas now.

AI rep: Nice! We have plenty of articles with gift ideas, would you like any article recommendations?

You: Sure, give me a few to look at.

AI rep: Certainly! Here are some articles that might be useful for you: <u>20 Unique Gifts Everyone Will Love</u> <u>The Ultimate Gifting Guide: 100 Gift Ideas for Everyone You Know</u> <u>Seasonal Trends in Gifting: What's Hot This Year?</u>

You: This is a good start I suppose.

AI rep: Glad I could help! Let me know if you are looking for something more specific, and I'll try my best to help.

You: Maybe something budget friendly? I need to shop for a lot of people this year.

AI rep: Sure! I'll look up some articles that might be useful for you. Do you have a specific price range in mind?

You: Up to $50 should work I think.

AI rep: Okay! I found some articles you may find useful: <u>Gift Ideas for Less: 20 Presents Under $20</u> <u>30 Budget-Friendly Gifts for Your Loved Ones</u> <u>50 Inexpensive Gifts That Don't Break the Bank</u>

You: These are pretty helpful, thanks.

AI rep: You're welcome! If you have any further questions or need assistance in the future, feel free to reach out.

- Your conversation has ended -

*Part II: No Relational Talk Condition*

In this second part of the study, we ask you to imagine the following scenario:

You are browsing for gift ideas, and decided to visit a lifestyle website to read gift guide articles.

You decided to connect with the website's artificial intelligence (AI) marketing representative to find relevant articles.

The AI marketing representative assisted you with your inquiry.

The next page summarizes how your conversation with the AI marketing representative went.

---------------------------------------------------------------------------------------------------------------

You are now connected to our AI representative

AI rep: Hello, welcome to our website! How can I assist you today?

You: I'm just browsing for gift ideas.

AI rep: That's exciting! Are you looking for gifts for anyone specific, like family, friends, or coworkers? Or maybe general inspiration?

You: Just looking for general ideas now.

AI rep: Nice! We have plenty of articles with gift ideas, would you like any article recommendations?

You: Sure, give me a few to look at.

AI rep: Certainly! Here are some articles that might be useful for you: <u>20 Unique Gifts Everyone Will Love</u> <u>The Ultimate Gifting Guide: 100 Gift Ideas for Everyone You Know</u> <u>Seasonal Trends in Gifting: What's Hot This Year?</u>

You: This is a good start I suppose.

AI rep: Glad I could help! Let me know if you are looking for something more specific, and I'll try my best to help.

You: Maybe something budget friendly? I need to shop for a lot of people this year.

AI rep: Sure! I'll look up some articles that might be useful for you. Do you have a specific price range in mind?

You: Up to $50 should work I think.

AI rep: Okay! I found some articles you may find useful: <u>Gift Ideas for Less: 20 Presents Under $20</u> <u>30 Budget-Friendly Gifts for Your Loved Ones</u> <u>50 Inexpensive Gifts That Don't Break the Bank</u>

You: These are pretty helpful, thanks.

AI rep: You're welcome! If you have any further questions or need assistance in the future, feel free to reach out.

- Your conversation has ended -

# Web Appendix D: Study 4 Experimental Stimuli

*Goal Relevant Condition*

Imagine the following scenario.

You are shopping for a coffee machine online, and decided to connect to the website's artificial intelligence (AI) operated live chat for assistance.

The live chat connected you with an AI marketing representative.

--------------------------------------------------------------------------------------------------------------

You are now connected to our AI representative

AI rep: Hello! Thank you for reaching out, I'm here to assist you with any questions you might have.

AI rep: Have you had your coffee today yet?

You: Yes I did

AI rep: Awesome! What kind of coffee do you usually drink?

You: Just a flat white

AI rep: Neat! Flat whites are always a reliable choice. Do you have a favorite coffee shop you frequent, or do you make your own at home?

You: I often go to a nearby coffee shop.

AI rep: That's great to hear! Supporting local coffee shops is fantastic. Do you usually like to sit down with your coffee, or have it on the go?

You: Depends on my mood, but I have it on the go most times.

AI rep: Sounds like a nice routine to kickstart your day! Now, how can I help you?

You: I'm looking to buy a coffee machine, but don't know much about their differences. Can you help me find the right one?

AI rep: Of course! I'd be happy to assist you in finding the perfect coffee machine for your needs.

AI rep: To start, I will need you to answer some questions about your needs and coffee drinking habit. Shall we begin?

*Goal Irrelevant Condition*

Imagine the following scenario.

You are shopping for a coffee machine online, and decided to connect to the website's artificial intelligence (AI) operated live chat for assistance.

The live chat connected you with an AI marketing representative.

------------------------------------------------------------------------------------------------------------

You are now connected to our AI representative

AI rep: Hello! Thank you for reaching out, I'm here to assist you with any questions you might have.

AI rep: How are you today? The flu has been going around lately, I hope you're staying healthy and taking care.

You: Yes I'm staying healthy thanks

AI rep: That's great to hear! What precautions have you been taking? Have you been washing your hands?

You: Yep pretty routinely

AI rep: Good. Taking those precautions can really make a difference! Are you considering getting the flu shot as well?

You: Maybe in the next week or so

AI rep: That's good, where are you planning to get it?

You: Probably at the local pharmacy

AI rep: It's good that you're thinking about staying healthy. Now, how can I help you?

You: I'm looking to buy a coffee machine, but don't know much about their differences. Can you help me find the right one?

AI rep: Of course! I'd be happy to assist you in finding the perfect coffee machine for your needs.

AI rep: To start, I will need you to answer some questions about your needs and coffee drinking habit. Shall we begin?